\documentclass[a4paper]{jpconf}
\usepackage{graphicx}
\begin{document}

\title{Strange stars at finite temperature}
\author{Subharthi Ray$^1$, Manjari Bagchi$^2$, Jishnu Dey$^2$ \& Mira Dey$^2$}

\address{$^1$ Inter University Centre for Astronomy and Astrophysics, Ganeshkhind, Pune, 411007, India}
\address{$^2$ Department of Physics, Presidency College, Kolkata 700073, India}


\begin{abstract}

We calculate strange star properties, using large $N_c$
approximation with built-in chiral symmetry restoration (CSM). We
used a relativistic Hartree Fock mean field approximation method,
using a modified Richardson potential with two scale parameters
$\Lambda$ and $\Lambda^{\prime}$, to find a new set of equation
of state (EOS) for strange quark matter. We take the effect of
temperature (T) on gluon mass, in addition to the usual density
dependence, and find that the transition T from hadronic matter
to strange matter is 80 MeV. Therefore formation of strange stars
may be the only signal for formation of QGP with asymptotic
freedom (AF) and CSM.
\end{abstract}

\section{Introduction}

There have been some exciting developments recently since the
four groups BRAHMS, PHENIX, PHOBOS and  STAR have analyzed RHIC
data.  These four groups reporting on RHIC, with gold on gold,
show that quark gluon plasma with asymptotic freedom and chiral
symmetry restoration may never be realized in RHIC although a
non-hadronic phase is reached.  Unfortunately, their conclusions
are negative in so far as finding of asymptotically free chirally
symmetric QCD state is not possible in these reactions - although
a new phase is formed which is quite distinct from the hadronic
phase. This phase is strongly interacting and is not fully
understood theoretically but it is certainly not QGP with
asymptotic freedom and chiral symmetry restoration.

Two decades back Witten\cite{wit} has proposed the existence of
strange matter and strange stars, and even today it is still
difficult to prove or disprove the existence of a state of
strange quark matter in its purest form. In literature, there are
several  EOSs for strange quark matter, starting from the Bag
model \cite{afo86,han86,ket95} to the recent models like the mean
field model with interacting quarks\cite{D98}, the perturbative
QCD approach\cite{fps01}, chiral chromodielectric
model\cite{mft03}, Dyson-Schwinger model\cite{blas99}, etc.
Subsequently, Rajagopal and Wilczek combined asymptotic freedom
and BCS theory to arrive at the color-flavor locked state of
quark matter\cite{rw00}, and ever since, a lot of studies on this
state and their application to quark matter EOSs are
made\cite{abr01}. Here, we developed a set of new EOSs using a two parameter
interaction potential.

\section{The Model}

The original  qq potential of Richardson\cite{rich} was designed to
obtain the mass spectrum of heavy mesons (Charmonium and
Upsilon). It takes care of two features of qq force, AF and
confinement, however, with the same scale, ${\Lambda}$ :

\begin{equation}
V_{ij} = \frac{12 \pi}{27}\frac{1}{ln(1 +  {({\bf k}_i- {\bf
k}_j)}^2 /\Lambda ^2)}\frac{1}{{({\bf k}_i - {\bf k}_j)}^2}
\label{eq:V}
\end{equation}
with ${\Lambda}$=400 MeV. It was further applied to light meson
spectroscopy  and baryon properties with the same value of
${\Lambda}$\cite{crater,ddt}. In strange stars,
the AF part is important and there ${\Lambda}$ would be much
smaller, around 100 MeV which is the scale for asymptotic
freedom as obtained from perturbative QCD. Indeed, for the self
bound high density strange quark matter Dey et  al.\cite{D98}
found that ${\Lambda}$ needed to be $\sim 100 $ MeV. Quarks are
deconfined at high density, as the bare potential is screened.
Confinement is softened and the AF part takes over. This bare
potential in a medium will be screened due to gluon propagation.
The temperature dependence of the screening in quark matter is
taken from Alexanian \& Nair \cite{alexnair}. The inverse Debye
screening length (the gluon mass) becomes :

\begin{equation}
(D^{-1})^2 = \frac{ 2 \alpha_0}{\pi} \sum_{i=u,d,s,}k^f_i
\sqrt{(k^f_i)^2 + m_i^2} ~+ 7.14 ~\alpha_0 ~T
\label{eq:gmt}
\end{equation}
where $k^f_i$, the Fermi momentum of the {\it i-th} quark is
obtained from the corresponding number density ${k^f_i} = (n _i
\pi^2)^{1/3}$ and $\alpha_0$ is the perturbative quark gluon
coupling.

In our model, chiral symmetry restoration at high density is
incorporated by assuming that the quark masses are density
dependent :
\begin{equation}
M_i = m_i + M_Q sech\left( \frac{n_B}{N n _0}\right), \;\;~~~ i =
u, d, s.
\label{eq:qm}
\end{equation}
where $n_B = (n_u+n_d+n_s)/3$ is the baryon number density; $n_0
= 0.17~fm^{-3}$ is the normal nuclear matter density; $n_u$,
$n_d$, $n_s$ are number densities of u, d and s quarks
respectively and $N$ is a parameter. The current quark masses
($m_i$) are taken as : $m_u = 4 \;MeV,\; m_d = 7 \;MeV,\; m_s =
150 \;MeV$. It is ensured that in strange matter, the chemical
potentials of the quarks satisfy ${\beta}$ equilibrium and charge
neutrality conditions. The parameters $M_Q$ and $N$ are adjusted
in such a way that the minimum value of $E/A$ for u,d,s quark
matter is less than that of Fe$^{56}$, so that u,d,s quark matter
can constitute stable stars. The minimum value of $E/A$ is
obtained at the star surface where the pressure is zero. The
surface is sharp since strong interaction dictates the
deconfinement point. However, the minimum value of $E/A$
for u-d quark matter is greater than that of Fe$^{56}$ so that
Fe$^{56}$ remains the most stable element in the non-strange world.

Recently the magnetic moments of $\Delta^{++}$ and $\Omega ^-$
with three u and three s valence quarks respectively have been
found in accurate experiments. These sensitive properties are
fitted with a two parameter modified Richardson potential with
different scales for confinement ($\sim 350$ MeV) and AF ($\sim
100 $ MeV)\cite{mmsj}. The baryonic properties depend more on the
confinement part and less on the AF part as baryons are confined
quark systems. It is natural to apply this new potential to star
calculation since they are constrained by baryonic data.

In our present calculation, we have modified the Richardson
potential as :

\begin{eqnarray}
V_{ij} = \frac{12 \pi}{27}\left[(\frac{1}{{\rm ln}(1 +
\frac{{({\bf k}_i- {\bf k}_j)}^2}{\Lambda
^2})-\frac{\Lambda^2}{({\bf k}_i- {\bf k}_j)^2}})
+\frac{{\Lambda^\prime}^2}{({{\bf k}_i- {\bf k}_j})^2}\right]
\times \frac{1}{({\bf k}_i - {\bf k}_j)^2}
\label{eq:Vtl}
\end{eqnarray}
with ${\Lambda}^{ \prime}$ taking care of the confinement property
and  ${\Lambda}$ that of asymptotic freedom.

The term $\left(\frac{1}{Q^2ln(1 +  Q^2 /\Lambda
^2)}-\frac{{\Lambda}^2}{Q^4}\right)$ is asymptotically zero for
large momentum transfer $Q^2 = ({{\bf k}_i- {\bf k}_j})^2$ and
the term $\frac{{\Lambda^\prime}^2}{Q^4}$ explains the
confinement reducing to a linear confinement for small $Q^2$.
The  appropriate values of ${\Lambda}$ and ${\Lambda}^{\prime}$
as obtained from a fit to baryonic properties
calculations\cite{mmsj} are $\Lambda \sim 100~MeV$ and
$\Lambda^\prime \sim 350~MeV$, which we used in the present
calculation.

Finite temperature T is incorporated in the system through the
Fermi function\cite{fintemp}:
\begin{equation}
FM(k,T)=\frac{1}{e^{(\epsilon - \epsilon_F)/T} + 1}
\end{equation}
with the flavour dependent single particle energy
\begin{equation}
\epsilon_i=\sqrt{k^2+M_i(\rho)^2}+U_i(k).
\label{eq:ep}
\end{equation}

\begin{equation}
I=\frac{\gamma}{2\pi^2}\int_0^\infty \phi(\epsilon)k^2FM(k,T) dk
\label{eq:i}
\end{equation}

I = number density for $\phi(\epsilon) = 1 $ and I = energy
density for $\phi(\epsilon) = \epsilon  $. $\gamma$ is the
spin-colour degeneracy, equal to 6.

The system is highly degenerate even at very high $T$ which is
around 80 $MeV$ since the chemical potential is very high, of
order several hundred $MeV$. The entropy is calculated as follows:
\begin{eqnarray}
S(T)=\int_0^\infty k^2[FM(k,T)ln(FM(k,T))  +
(1-FM(k,T))ln(1-FM(k,T))] dk
\label{eq:s}
\end{eqnarray}
The pressure is calculated from the free energy $f = \epsilon -
Ts$ as follows:
\begin{equation}
P=\sum_i\rho_i\frac{\partial f_i}{\partial \rho_i} - f_i
\label{pressure}
\end{equation}
With the obtained EOS (pressure vs density), we solve the Tolman
Oppenheimer Volkov equation, to get the stellar structure.

\begin{figure}
\begin{minipage}[t]{8cm}
\begin{center}
\includegraphics[width=7.5cm,clip]{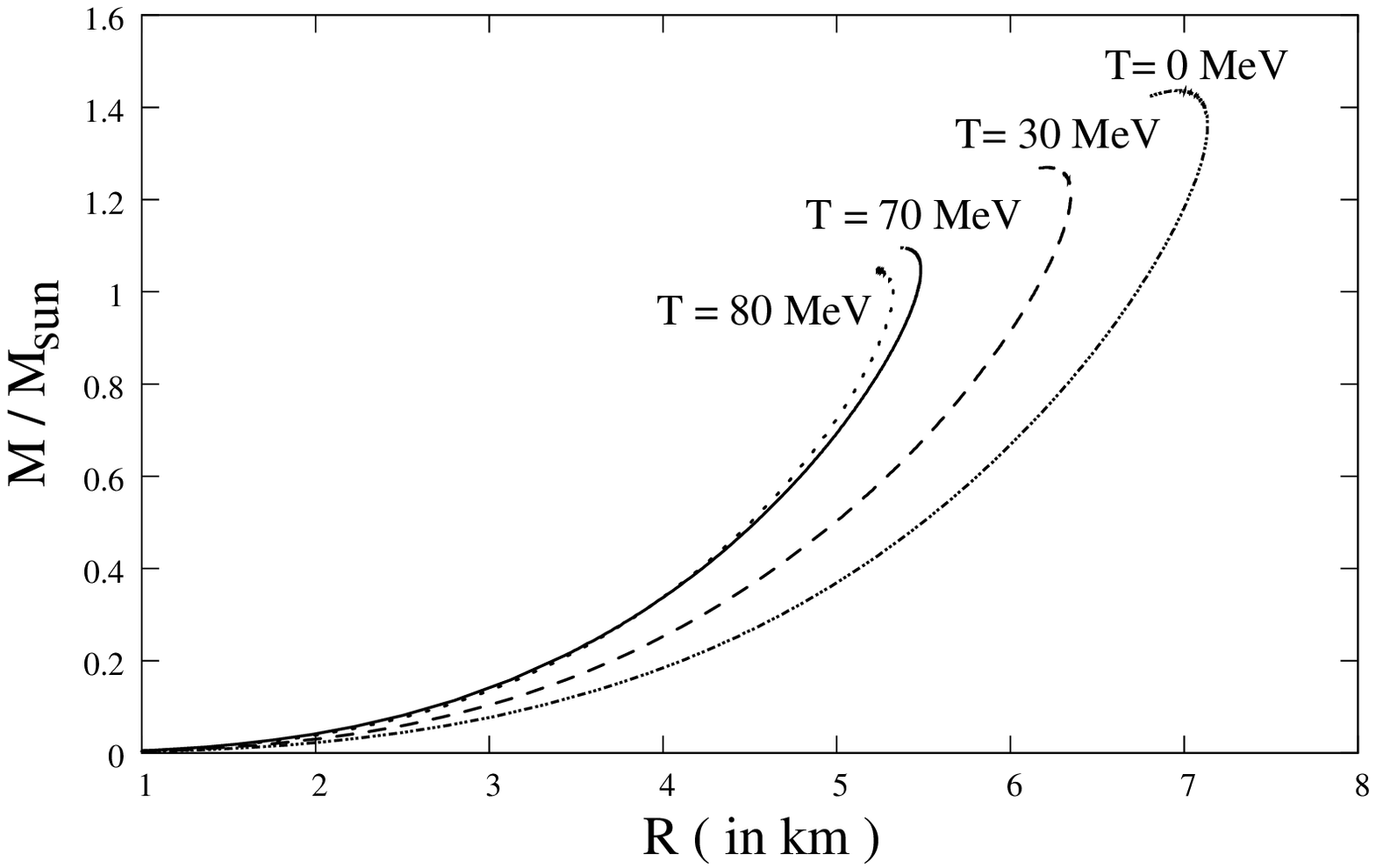}
\caption{\label{m-r} Mass Radius relations at different temperatures, for
$\Lambda=100~MeV$ and $\Lambda^\prime=350~MeV$, $N=3$ and
$\alpha_0=.2$. Although the maximum mass limit decreases, but for
a particular radius, we see the mass to be high for higher
temperature. }
\end{center}
\end{minipage}
\hfill
\begin{minipage}[t]{7.5cm}
\begin{center}
\includegraphics[width=7.5cm,clip]{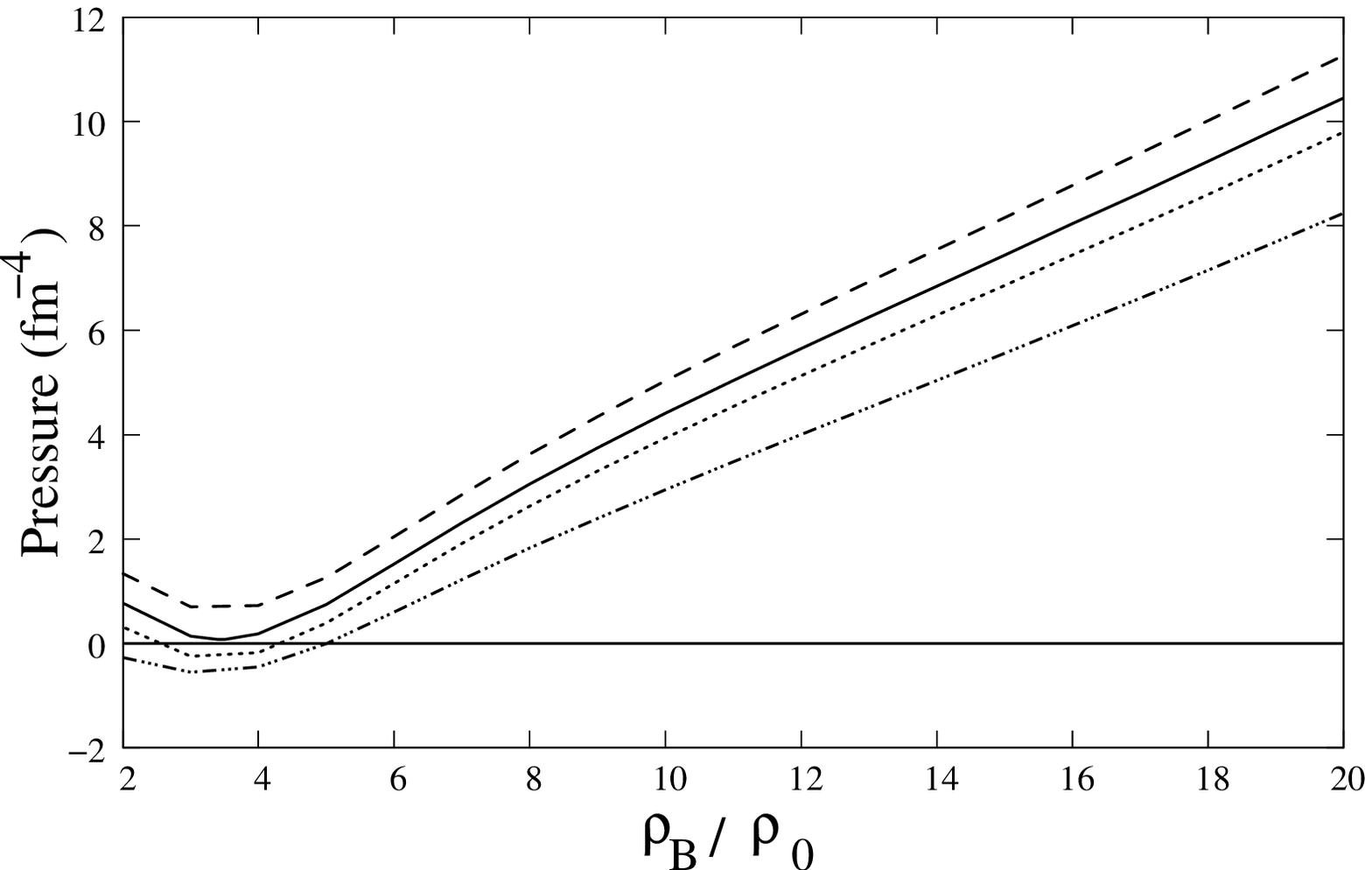}
\caption{\label{eos}EOS for the set of parameters $\Lambda=100~MeV$ and $\Lambda^\prime=350~MeV$, $N=3$ and
$\alpha_0=.2$ and for temperatures 0, 50, 80 and 90 Mev (bottom to
top). We see that for 90 MeV, there is no stable configuration.}
\end{center}
\end{minipage}
\end{figure}

\begin{figure}
\begin{minipage}[t]{8cm}
\begin{center}
\includegraphics[width=7.5cm,clip]{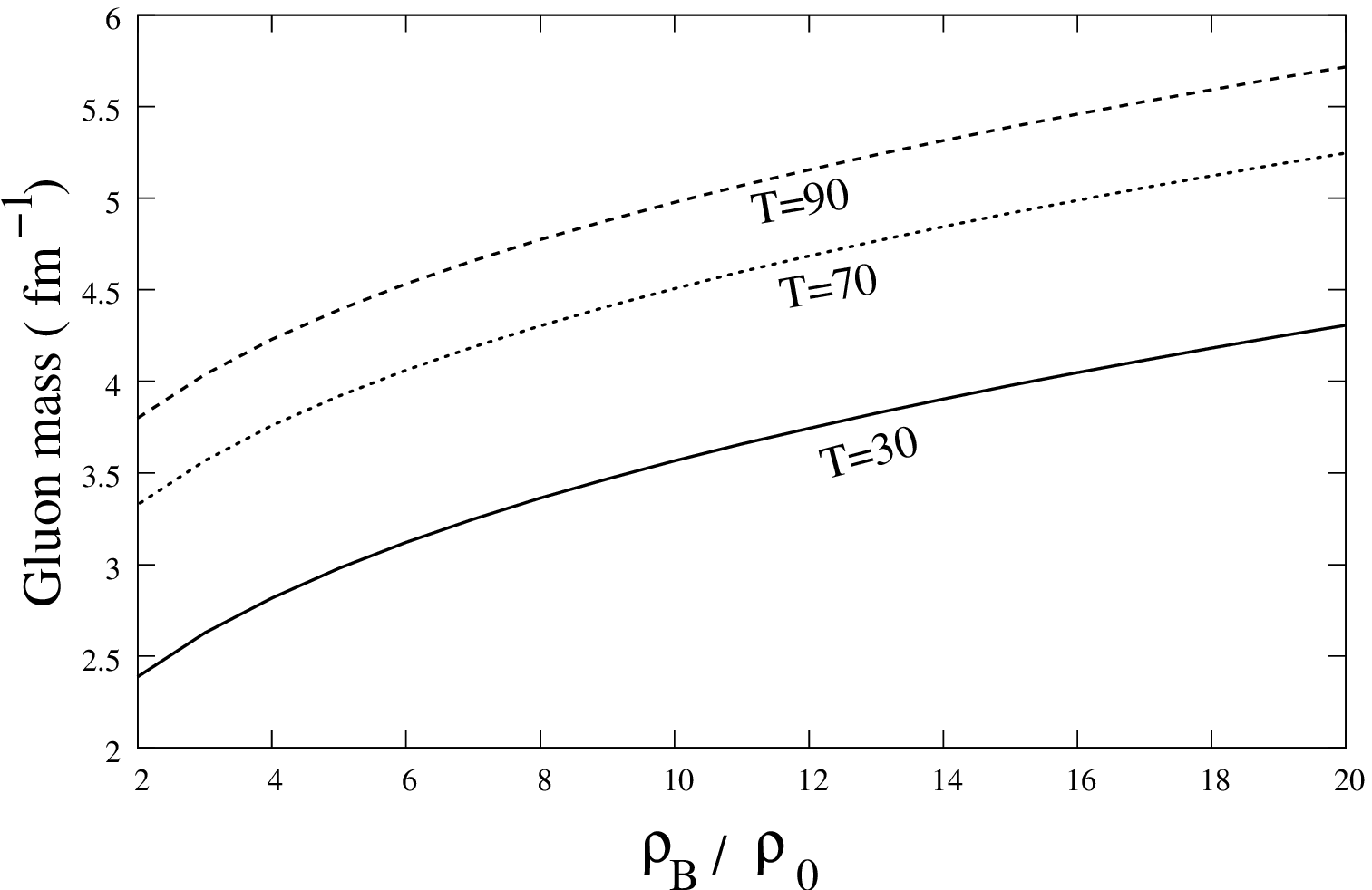}
\caption{\label{gmasstemp}Mass of the gluon (inverse Debye screening length)
increases with density and also with temperature.}
\end{center}
\end{minipage}
\hfill
\begin{minipage}[t]{7.5cm}
\begin{center}
\includegraphics[width=7.5cm,clip]{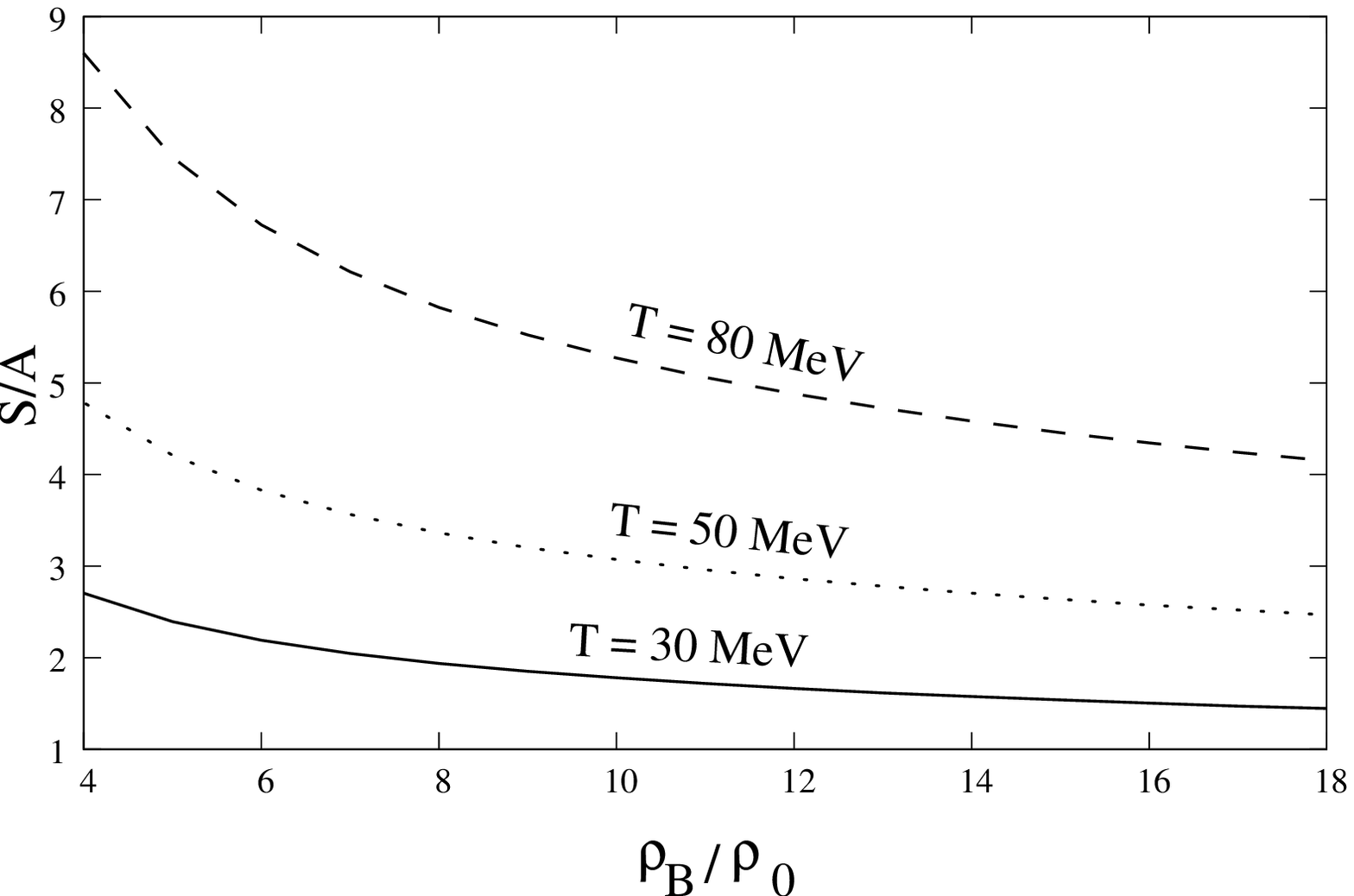}
\caption{\label{sbya}The entropy per baryon decrease with density. In the nuclear matter
limit (say 1.5$\times$ normal nuclear matter density), it is found
to match exactly with the experimental results}
\end{center}
\end{minipage}
\end{figure}

\section{Conclusions and summary:}
A set of new EOS for strange matter is presented, using the
Richardson potential with the value of ${\Lambda}^{\prime}~
\simeq~300~{\rm to}~350~MeV$ and the value of ${\Lambda}~=
~100~MeV$, the two scales for confinement and AF respectively.
This is then a good inter-quark potential as it also explains
both the properties of the deconfined quark matter (our present
work) and the properties of confined 3u and 3s baryons\cite{mmsj}.
Although the confinement part ${\Lambda}^{\prime}~$ is stronger,
leading to a sharp surface, - it is softened by medium effect,
developing a screening length. Inside the star, the AF part is
more important. We have found that for a wide range of parametric
variation, the strange matter EOS gives minimum energy of E/A,
much less than compared to ${(E/A)}_{Fe^{56}}=~930.6~MeV$ which
ensures that the system is absolutely stable, and unlike neutron
star like structure, it is not gravitational force alone that
binds the system. However, for this present article, we have
presented with only one such EOS. Considering temperature
dependent screening of the potential, we have found that strange
stars can sustain stable configurations up to a temperature of 80
$MeV$; this value of the temperature is very close to Witten's
scenario of cosmic phase separation.




\end{document}